\documentstyle[12pt]{article}
\newcommand\be{\begin{equation}}
\newcommand\ee{\end{equation}}
\newcommand\bea{\begin{eqnarray}}
\newcommand\eea{\end{eqnarray}}
\newcommand\ket[1]{|#1\rangle}

\newcommand{\fatalpha}{{\bf \alpha \kern -0.44em \alpha}}
\newcommand{\fatsigma}{{\bf \sigma \kern -0.54em \sigma}}
\newcommand{\tpchi}{{\bf \chi \kern -0.35em \chi}}
\newcommand{\llambda}{{\bf \lambda \kern -0.45em \lambda}}

\bibliography{plain}
\title{\bf Quantum state transfer in atom-cavity systems with uncolored Cayley interacting networks}\vspace{20mm}
\author{ R. Sufiani$^{a,b}$
  \thanks{E-mail:sofiani@tabrizu.ac.ir}
 \\ $^a$ {\small Department of Theoretical Physics and Astrophysics,
University of Tabriz, Tabriz 51664, Iran.} \\ $^b$ {\small Institute
for Studies in Theoretical Physics and Mathematics, Tehran
19395-1795, Iran.}} \pagebreak

\vspace{20mm}
\begin{document}
\maketitle \vspace{15mm}
\newpage
\begin{abstract}
Considering the two-photon exchange interaction between $n$ coupled
cavities each of them containing a two level atom, the atomic and
photonic state transfer is investigated. In fact, $n$ atom-cavity
systems are considered to be distributed on the nodes of an
uncolored Cayley network and interact with each other via the
adjacency matrix of the corresponding network. Then, by employing
the photon-excitation conservation and also the algebraic structure
of the networks, such as irreducible characters of the groups
associated with the networks, some suitable basis for the
atom-cavity state space is introduced based on the corresponding
\textit{generalized} Fourier transform, so that the Hamiltonian of
the whole system, is block diagonalized with two-dimensional blocks.
Then, by solving the corresponding Schr\"{o}dinger equation exactly,
quantum state transfer and also entanglement generation between the
atoms or the photons are discussed. For instance, the probability
amplitudes associated with the photon transition between the
cavities or excitation transition between the atoms are obtained in
terms of the irreducibles characters of the corresponding network
and the hopping parameter $\xi$ between the cavities.

 {\bf Keywords: coupled cavities, two-photon exchange, hopping strength, two-level atoms, generation of entanglement, excitation and photon transfer, uncolored Cayley graphs, Generalized Fourier
 transform}\\

{\bf PACs Index: 03.65.Ud }
\end{abstract}

\vspace{70mm}
\newpage
\section{Introduction}
In the past years, several schemes for quantum information
transition and the generation and distribution of entanglement have
been designed and implemented in a number of physical systems (see
for example \cite{11}-\cite{20}). Atoms and ions are particularly
considered as tools for storing quantum information in their
internal states. On the other hand, photons represent the best qubit
carrier for the fast and reliable communication over long distances
\cite{ref5,ref6}. Recently, using photons in order to achieve
efficient quantum information transmission between spatially distant
atoms has investigated in several works \cite{Alex1, ref1, ref,
Alex2, ozgur, suf}. The main idea, is to utilize two photon exchange
hopping between optical cavities containing the atoms. One of the
known models in quantum optics describing the atom-field interaction
is the Jaynes-Cummings Hamiltonian \cite{book1, book2}. In Refs.
\cite{Alex1, Alex2}, entanglement properties of two and three
atom-cavities coupled via two-photon exchange interaction, was
studied in detail. The motivation of interest to such systems is
their ability to be used in quantum switching, quantum communication
and computation, and quantum phase transition applications.

In the recent work \cite{suf}, the approach of Alexanian, et. al has
generalized to the case of $n$ atom-cavities - instead of two or
three cavities- all of them interacting with each other by hopping
strength $\xi$ (in that case the cavities were distributed on a
complete network $K_n$ at which all of the vertices connected to
each other). In this paper, I consider that the atom-cavity systems
are interacting with each other via the more general connectivity
networks (graphs) called uncolored Cayley networks. Due to the fact
that these graphs are constructed via finite groups, one can
naturally use their algebraic structure in order to state transfer
or entanglement generation. In fact, I consider the Cayley graphs
associated with abelian groups and employ the group theoretical
properties of these networks such as generalized Fourier transform
associated with the corresponding groups (the unitary transform
which diagonalizes the regular representation of the groups). Then,
by using the photon-excitation conservation symmetry of the
Hamiltonian, some suitable basis for the atom-cavity state space is
introduced, so that the Hamiltonian of the whole system, is block
diagonalized with two dimensional blocks. Then, by solving the
corresponding Schr\"{o}dinger equation exactly, state transfer
between atoms or photons is discussed. For instance, the probability
amplitudes associated with the photon transition between the
cavities or excitation transition between the atoms are obtained in
terms of the irreducibles characters of the groups associated with
the corresponding networks, and the hopping parameter $\xi$ between
the cavities.

The paper is organized as follows. In section 2, a preliminary is
given in order to recall the definition of uncolored Cayley
networks. In section $3$, the model describing the system of $n$
identical atom-cavities is introduced. The block-diagonalization of
the Hamiltonian of the system, solving the corresponding
Shr\"{o}dinger equation for time dependent probability amplitudes of
the state of the system, and entanglement generation between atoms
or photons are discussed in this section. Section $4$ is devoted to
discussions about state transfer (excitation or photon transfer
between the atoms or the cavities). In sections $5$, two examples of
Cayley networks (Cycle network $C_n$ and the hypercube network
$Q_d$) are discussed in detail, in order to clarify the results of
the paper. The paper is ended with a brief conclusion.
\section{Preliminary: Uncolored Cayley networks}
For a given finite group $G$, we have always a set of group elements
$S$ so that all of the group elements can be generated via the
possible multiplications of the elements of $S$ and their powers.
The set $S$ is called the generating set of $G$. The Cayley graph
$\Gamma (G,S)$ is a colored directed graph constructed as follows:
Each element $g$ of $G$ is assigned a vertex: the vertex set
$V(\Gamma)$ of $\Gamma$ is then identified with $G$. Each generator
$s$ of $S$ is assigned a color $c_s$. For any $g\in G$ and $s\in S$,
the vertices corresponding to the elements $g$ and $gs$ are joined
by a directed edge of color $c_s$. Thus the edge set $E(\Gamma)$
consists of pairs of the form $(g,gs)$ with $s\in S$ providing the
color.

The set $S$  is usually assumed to be finite, symmetric (i.e.
$S=S^{-1}$ ) and not containing the identity element of the group.
In this case, the uncolored Cayley graph is an ordinary graph: its
edges are not oriented and it does not contain loops. Then, the
adjacency matrix $A$ corresponding to the Cayley graph $\Gamma
(G,S)$ is defined as:
\begin{equation}\label{relation1}A=\sum_{s\in S}R(s),\end{equation}
where $R(s)$ is the regular representation of the group element
$s\in S$.

We will assume the Cayley graphs corresponding to the Abelian group,
so that the adjacency matrix $A$ can be diagonalized via generalized
Fourier transform $P$ defined by
$P_{ij}=\frac{1}{\sqrt{n}}\chi_{i}(j)$ where, $\chi_i(j)$, for
$i,j=1,2,\ldots, n \;(=|G|)$, is the $i$-th irreducible character of
the $j$-th group element, and $|G|$ is the cardinality of the group.
Then, the eigenvalues of the adjacency matrices $A_{k}$ are given by
\begin{equation}\label{eqx}x_{i}=\sum_{s\in S}\chi_{i}(s).
\end{equation}
It could be noticed that, from the orthogonality relation for the
irreducible characters \cite{Gordon},
\begin{equation}\label{ort.}
 \frac{1}{n}\sum_k\chi_i(k)\chi^*_j(k)=\delta_{ij},
\end{equation}
one can see that $(P^{-1})_{ij}=\frac{1}{\sqrt{n}}\chi^*_{j}(i)$ and
so we have $P^{-1}=P^{\dag}$.
\section{The Model: $n$ coupled cavities via two-photon exchange interaction}
We will consider $n$ identical cavities each containing a two-level
atom, where the cavities are coupled via two photon hopping between
them. In fact, we consider that the cavities are located at the
nodes of an uncolored Cayley network with $n$ nodes and each cavity
interacts with all of the adjacent cavities, via two-photon
exchange.

Let us first introduce the Hamiltonian of one individual atom-cavity
as a two-photon JCM \cite{JC} (assume $\hbar=1$):
\begin{equation}\label{eq0}
H^{(i)}=\omega_a s^{(i)}_z+\omega_c
a^{\dag}_ia_i+\lambda(\sigma^{(i)}_{eg}a^2_i+\sigma^{(i)}_{ge}a^{\dag
2}_i),
\end{equation}
where, the operators $a_i$ and $a_i^{\dag}$ are photonic
annihilation and creation operators of the $i$-th cavity,
$\sigma^{(i)}_{ab}=|a\rangle^{{{(i)}{(i)}}}\langle b|$, for $ i = 1,
2,\ldots, n$ denote the atomic transition operators for the $i$-th
cavity referring to either the ground (g) or excited (e) state;
$\omega_a$ is the energy separation between two levels $|g\rangle$
and $|e\rangle$,
$s^{(i)}_z=\frac{1}{2}\sigma^{(i)}_z=\frac{1}{2}(|e\rangle^{{{(i)}{(i)}}}\langle
e|-|g\rangle^{{{(i)}{(i)}}}\langle g|)$, and $\omega_c$ is the
resonance frequency of the cavity.

In the Hamiltonian (\ref{eq0}), the excitation-number operator
$$\hat{N}^{(i)}=a_i^{\dag}a_i+s^{(i)}_z$$
is a constant of motion for the $i$-th atom-cavity subsystem, i.e,
we have $[H^{(i)}, \hat{N}^{(i)}]=0$ for each $i=1,2,\ldots, n$.
Now, the Hamiltonian for the $n$ cavities is given by:
\begin{equation}\label{H}
H=\sum_{i=1}^n H^{(i)}+\xi \sum_{i,j=1; i\sim
j}^n(a_i^{2{\dag}}a_j^2+a_j^{2{\dag}}a_i^2),\end{equation} where,
$i\sim j$ means that the cavities $i$ and $j$ are adjacent on the
network. The last term in the Hamiltonian (\ref{H}) is the
two-photon exchange interaction between the cavities, characterized
by the hopping rate $\xi$.

The operator $\hat{N}=\sum_{i=1}^n {\hat{N}}^{(i)}$ commutes with
the Hamiltonian (\ref{H}) and so we can reduce the Hamiltonian to
the subspace spanned with the eigenstates of $\hat{N}$ and consider
the time evolution of the states in this subspace. For a given
eigenspace of $\hat{N}$ with eigenvalue $N$, the maximum possible
number of photons in a cavity is $N$ when the corresponding atom is
in the ground state, which occurs when there are no photons present
in the other cavities and the atoms are also in the ground state.
Then, the total number of photons in the system will be $N$. The
constant number of total photons determines the subspace or the
manifold in which the states evolve in time (the initial state of
the system determines the constant number $N$). For simplicity, we
will consider the manifold with $N=\frac{2-n}{2}$. In this case,
each single atom-cavity system can take one of the three possible
states $\ket{g,0}$, $\ket{g,2}$ or $\ket{e,0}$, and so, the total
possible states that the system of $n$-cavities can take, are $3^n$
states. Due to the consistency of total $N=\frac{2-n}{2}$, the only
possible states which we can have, are $2n$ states instead of $3^n$
ones. In fact, these $2n$ states are eigenstates of $\hat{N}$ with
eigenvalue $2$, and the $3^n$-dimensional Hamiltonian $H$ is reduced
to $2n$-dimensional one in the bases which span the eigenspace of
$\hat{N}$ with the corresponding eigenvalue $2$. The bases states
that span this subspace or manifold, are given by:
$$\ket{c_i}=\ket{g,0}\ldots \ket{g,0}\underbrace{\ket{g,2}}_{i-th}\ket{g,0}\ldots \ket{g,0},$$
$$\ket{a_i}=\ket{g,0}\ldots \ket{g,0}\underbrace{\ket{e,0}}_{i-th}\ket{g,0}\ldots \ket{g,0},$$
for $i=0,1,\ldots, n-1$. Indeed, these bases span the eigenspace of
$\hat{N}$ with eigenvalue $2$, i.e., we have $\hat{N} (\alpha
\ket{c_i}+\beta \ket{a_i})=\frac{2-n}{2}(\alpha \ket{c_i}+\beta
\ket{a_i})$. Therefore, the general time dependent state of the
$n$-cavity system is given by
\begin{equation}\label{sai}
\ket{\psi(t)}=\sum_{i=0}^{n-1}(C_i(t)\ket{c_i}+A_i(t)\ket{a_i}).
\end{equation}
Then, one can easily show that
$$\sum_{i=1}^n H^{(i)}|a_k\rangle=\frac{2-n}{2}\omega_a |a_k\rangle+\sqrt{2}\lambda|c_k\rangle,$$
$$\sum_{i=1}^n H^{(i)}|c_k\rangle=\frac{4\omega_c-n\omega_a}{2}|c_k\rangle+\sqrt{2}\lambda|a_k\rangle,$$
$$H_{_{int.}}|c_k\rangle\equiv \xi \sum_{i,j=1; i\sim
j}^n(a_i^{2{\dag}}a_j^2+a_j^{2{\dag}}a_i^2)|c_k\rangle=2\xi
\sum_{i\sim k} |c_i\rangle,\;\;\;\ H_{_{int.}}|a_k\rangle=0; \;\;\;\
k=0,1,\ldots, n-1.$$ Now, by considering the order of bases as
$\ket{c_0}, \ket{a_0},\ldots, \ket{c_{n-1}}, \ket{a_{n-1}}$, and
defining $\Delta=\omega_a-2\omega_c$, the Hamiltonian $H$ takes the
following direct product form
\begin{equation}\label{H1}H=I_n \otimes \left(\begin{array}{cc}
              \omega+\Delta/2 & \sqrt{2}\lambda \\
                  \sqrt{2}\lambda & \omega-\Delta/2 \\
                \end{array}\right)+2\xi A\otimes \left(\begin{array}{cc}
              1 & 0 \\
                 0 & 0 \\
                \end{array}\right),
                \end{equation}
                where, $\omega:=\frac{1}{2}[(2-n)\omega_a-\Delta]$ and $A$ is the adjacency matrix of the corresponding network containing the atom-cavities on its nodes.
We will assume that the cavities are interacting with each others
via an uncolored Cayley network as connectivity network, so that
$A\equiv A_1$ is given by (\ref{relation1}). That is, the photons
only hope between the cavities which are adjacent in the network.

Now, due to the fact that the adjacency matrix $A$ is diagonalized
by the generalized Fourier transform $P_{kl}:=
\frac{1}{\sqrt{n}}\chi_k(\alpha_l)$, one can naturally consider the
new transformed bases $\{\ket{c_j}', \ket{a_j}'\}_{j=0}^{n-1}$ as:
$$
\ket{c_j}':=
\frac{1}{\sqrt{n}}\sum_{i=0}^{n-1}\chi_j(\alpha_i)\ket{c_i},$$
\begin{equation}\label{basis}
\ket{a_j}':=
\frac{1}{\sqrt{n}}\sum_{i=0}^{n-1}\chi_j(\alpha_i)\ket{a_i}.
\end{equation}
Then, by considering the ordering $\{\ket{c_0}', \ket{a_0}';
\ldots;\ket{c_{n-1}}', \ket{a_{n-1}}'\}$, the Hamiltonian (\ref{H1})
takes the following block diagonalized form:
\begin{equation}\label{H2}H=I_n \otimes \left(\begin{array}{cc}
              \omega+\Delta/2 & \sqrt{2}\lambda \\
                  \sqrt{2}\lambda & \omega-\Delta/2 \\
                \end{array}\right)+2\xi\;\ diag{(x_0,x_1,\ldots,x_{n-1})}\otimes \left(\begin{array}{cc}
               1 & 0 \\
                 0 & 0 \\
                \end{array}\right),
                \end{equation}
where, $diag {(x_0,x_1,\ldots,x_{n-1})}$ is the $n\times n$ diagonal
matrix with diagonal entries as $x_i=\chi_i(\alpha_1)$ with
$i=0,1,\ldots, n-1$. Therefore, the $i$-th block of the Hamiltonian
is given by
\begin{equation}\label{H2''}H^{(i)}=\left(\begin{array}{cc}
              \omega+\Delta/2+2\xi x_i & \sqrt{2}\lambda \\
                  \sqrt{2}\lambda & \omega-\Delta/2 \\
                \end{array}\right).
                \end{equation}

with eigenvalues $E_{\pm}^{(i)}=(\omega+\xi x_i)\pm
\sqrt{2\lambda^2+(\frac{\Delta}{2}+\xi x_i)^2}$.
  Now,
by using the Schr\"{o}dinger equation of motion $i\hbar
\frac{\partial}{\partial t}\ket{\psi}=H\ket{\psi}$, the equations of
motion are given by:
$$i\dot{C'_i}=[\omega+\Delta/2+2\xi x_i]C'_i+\sqrt{2}\lambda A'_i;$$
\begin{equation}\label{eq.mo}
\hspace{-1.5cm}i\dot{A'_i}=\sqrt{2}\lambda C'_i+(\omega-\Delta/2)
A'_i,
\end{equation}
for $ i=0,1,\ldots, n-1$. Solving the corresponding differential
equations, one can obtain
$$\hspace{-1cm}C'_i(t)=\frac{e^{-i(\omega+\xi x_i)t}}{\sqrt{2\lambda^2+(\frac{\Delta}{2}+\xi x_i)^2}}\{[\sqrt{2\lambda^2+(\frac{\Delta}{2}+\xi x_i)^2}\cos t\sqrt{2\lambda^2+(\frac{\Delta}{2}+\xi x_i)^2}-$$
$$i(\frac{\Delta}{2}+\xi x_i)\sin t\sqrt{2\lambda^2+(\frac{\Delta}{2}+\xi
x_i)^2}]C'_i(0)-i\sqrt{2}\lambda\sin
t\sqrt{2\lambda^2+(\frac{\Delta}{2}+\xi x_i)^2}A'_i(0)\},$$
$$\hspace{-1cm}A'_i(t)=\frac{e^{-i(\omega+\xi x_i)t}}{\sqrt{2\lambda^2+(\frac{\Delta}{2}+\xi x_i)^2}}\{[\sqrt{2\lambda^2+(\frac{\Delta}{2}+\xi x_i)^2}\cos t\sqrt{2\lambda^2+(\frac{\Delta}{2}+\xi x_i)^2}+$$
\begin{equation}\label{eq1}i(\frac{\Delta}{2}+\xi x_i)\sin t\sqrt{2\lambda^2+(\frac{\Delta}{2}+\xi
x_i)^2}]A'_i(0)-i\sqrt{2}\lambda\sin
t\sqrt{2\lambda^2+(\frac{\Delta}{2}+\xi x_i)^2}C'_i(0)\},
\end{equation}
In the resonance case $\Delta=0$, by considering
$\lambda=\frac{1}{\sqrt{2}}$, the coefficients (\ref{eq1}) read as
$$\hspace{-1cm}C'_i(t)=\frac{e^{-i(\omega+\xi x_i)t}}{\sqrt{1+(\xi x_i)^2}}\{[\sqrt{1+(\xi x_i)^2}\cos t\sqrt{1+(\xi x_i)^2}-i\xi x_i\sin t\sqrt{1+(\xi
x_i)^2}]C'_i(0)-i\sin t\sqrt{1+(\xi x_i)^2}A'_i(0)\},$$
\begin{equation}\label{eq1'0}
\hspace{-1cm}A'_i(t)=\frac{e^{-i(\omega+\xi x_i)t}}{\sqrt{1+(\xi
x_i)^2}}\{[\sqrt{1+(\xi x_i)^2}\cos t\sqrt{1+(\xi x_i)^2}+i\xi
x_i\sin t\sqrt{1+(\xi x_i)^2}]A'_i(0)-i\sin t\sqrt{1+(\xi
x_i)^2}C'_i(0)\}.
\end{equation}

By using (\ref{basis}), one can obtain the time dependence of the
coefficients $C_i(t)$ and $A_i(t)$ of the state of the system in
(\ref{sai}) via the inverse $P$ transform as,
$$C_i(t)=\frac{1}{\sqrt{n}}\sum_{j=0}^{n-1}\chi^{*}_j(\alpha_i)C'_j(t),$$
\begin{equation}\label{eq3}
A_i(t)=\frac{1}{\sqrt{n}}\sum_{j=0}^{n-1}\chi^{*}_j(\alpha_i)A'_j(t),
\end{equation}
where, $*$ denotes the complex conjugate. It should be pointed out
that, for the probabilities associated with the state of the system
as a superposition of atomic states $\ket{a_i}$, and that of
photonic states $\ket{c_i}$ (denoted by $P_a(t)$ and $P_c(t)$,
respectively), the equation (\ref{eq3}) gives
$$P_a=\sum_{i=0}^{n-1}|A_i(t)|^2=\frac{1}{n}\sum_{j,k}\underbrace{\sum_{i}\chi^{*}_j(\alpha_i)\chi_k(\alpha_i)}_{n\delta_{jk}} A'_j(t)A'_k(t)=\sum_{j=0}^{n-1}|A'_j(t)|^2$$
$$P_c=\sum_{i=0}^{n-1}|C_i(t)|^2=\frac{1}{n}\sum_{j,k}\underbrace{\sum_{i}\chi^{*}_j(\alpha_i)\chi_k(\alpha_i)}_{n\delta_{jk}} C'_j(t)C'_k(t)=\sum_{j=0}^{n-1}|C'_j(t)|^2$$
For instance, considering the initial state
$$\ket{\psi(0)}=\frac{1}{\sqrt{n}}(\ket{g,2}\ket{g,0}...
\ket{g,0}+\ket{g,0}\ket{g,2}\ket{g,0}...\ket{g,0}+\ldots+
\ket{g,0}...\ket{g,0}\ket{g,2})=$$
\begin{equation}\label{initial}\ket{gg\ldots g}\otimes
\frac{1}{\sqrt{n}}(\ket{200... 0}+\ket{020... 0}+\ldots+\ket{0...
002}),
\end{equation}
in which the atoms are initially unentangled, whereas the photons
are in a multipartite entangled $W$-state. Then, the initial
conditions are given by $A_j(0)=0$ and $C_j(0)=\frac{1}{\sqrt{n}}$
for all $j=0,1,..., n-1$, and we obtain via equation (\ref{basis}),
$A'_i(0)=0$, and $C'_i(0)=\frac{1}{n} \sum_{j}\chi_i(\alpha_j)$ for
all $i$. Then, with the aid of Eq. (\ref{eq1}), we obtain
$$P_a(t)=\sum_{i=0}^{n-1}|A'_i(t)|^2=\sum_i\sum_{j,k}\frac{2\lambda^2\sin^2t\sqrt{2\lambda^2+(\frac{\Delta}{2}+\xi
x_i)^2}}{n^2[2\lambda^2+(\frac{\Delta}{2}+\xi
x_i)^2]}\chi^{*}_i(\alpha_j)\chi_i(\alpha_k),$$
\begin{equation}\label{pr}
\hspace{-8cm}P_c(t)=1-P_a(t).
\end{equation}
The above result indicates that, in the limit of large
$\xi\rightarrow \infty$, we have $P_c(t)\simeq 1$ for every time
$t$, i.e., for large enough $\xi$, all of the atoms will be at their
ground state $\ket{g}$ at every time $t$.

In particular, at the resonance case $\Delta=0$, we obtain
$$P_a(t)=\sum_i\sum_{j,k}\frac{2\lambda^2\sin^2t\sqrt{2\lambda^2+\xi^2
x_i^2}}{n^2[2\lambda^2+\xi^2
x_i^2]}\chi^{*}_i(\alpha_j)\chi_i(\alpha_k),$$ where by taking
$\lambda=\frac{1}{\sqrt{2}}$, we have
$$P_a(t)=\sum_i\sum_{j,k}\frac{\sin^2t\sqrt{1+\xi^2
x_i^2}}{n^2[1+\xi^2 x_i^2]}\chi^{*}_i(\alpha_j)\chi_i(\alpha_k).$$

From the above result it can be seen that for small hopping strength
$\xi\ll$, we have
$$P_a(t)=\frac{\sin^2t}{n^2}\sum_{j,k}\underbrace{\sum_i\chi^{*}_i(\alpha_j)\chi_i(\alpha_k)}_{n\delta_{jk}}=\sin^2t,$$
which indicates that for times $T=\frac{(2k+1)\pi}{2}$, we have
$P_a(t)=1$, whereas for times $T=l\pi$, we have $P_c(t)=1$, for any
$k,l\in {Z}$. In other words, for small hopping $\xi$, the initial
state (\ref{initial}) evolves as
$$\ket{\psi(\frac{\pi}{2})}=(\frac{-i}{\sqrt{n}}\sum_{i=1}^n \ket{g\ldots g \underbrace{e}_{i} g\ldots g})\otimes \ket{00\ldots 0}$$
after time $T=\frac{\pi}{2}$ for which we have
$A_i(\frac{\pi}{2})=\frac{-i}{\sqrt{n}}$ for all $i$, and so the
initially unentangled atoms become entangled and the initial
photonic $W$-state oscillates between atomic and photonic $W$-states
periodically, with time period $\frac{\pi}{2}$. In order to evaluate
the amount of quantum entanglement between two atoms labeled by $l$
and $m$, one can calculate the state of the corresponding two atoms
as
$$\rho^{lm}_{_{at}}(\frac{\pi}{2})=Tr_{_{ph; k\neq l,m}}(\ket{\psi(\frac{\pi}{2})}\langle \psi(\frac{\pi}{2})|)=(1-\frac{2}{n})\ket{gg}\langle gg|+\frac{1}{n}(\ket{eg}\langle eg|+$$
$$\ket{ge}\langle ge|+\ket{eg}\langle ge|+\ket{ge}\langle eg|),$$
where the partial trace is taken over photonic states and also over
the atomic states corresponding to the atoms other than the atoms
$l$ and $m$. Now, the Peres-Horodecki criteria \cite{peres1, peres2}
known also as positive partial transpose (PPT) criteria can be
checked. The partial transpose of $\rho^{lm}_{_{at}}(\frac{\pi}{2})$
with respect to the first subsystem is given by
$$[\rho^{lm}_{_{at}}(\frac{\pi}{2})]^{T_1}=\frac{1}{n}\left(\begin{array}{cccc}
                                                         n-2 & 0 & 0 & 1 \\
                                                              0 & 1 & 0 & 0 \\
                                                              0 & 0 & 1 & 0 \\
                                                              1 & 0 & 0 & 0 \\
                                                            \end{array}\right),$$
                                                            so that
                                                            the
                                                            corresponding
                                                            eigenvalues
                                                            $\lambda_i$
                                                            are
                                                            given by
                                                            $\frac{1}{n}$
                                                            (with double degeneracy)
                                                            and $\lambda_{\pm}=\frac{1}{2n}(n-2\pm
                                                            \sqrt{(n-2)^2+4})$,
                                                            and the
                                                            negativity
                                                            is then
                                                            $N(\rho^{lm}_{_{at}}(\frac{\pi}{2}))=\sum_i|\lambda_i|-1=-2\lambda_-=\frac{1}{n}[
                                                            \sqrt{(n-2)^2+4}-(n-2)]$.
                                                            Then,
                                                            for the
                                                            case of
                                                            two
                                                            cavities,
                                                            $n=2$,
                                                            we have
                                                            $N(\rho_{_{at}}(\frac{\pi}{2}))=1$
                                                            and the
                                                            initially
                                                            unentangled
                                                            atoms,
                                                            become
                                                            maximally
                                                            entangled
                                                            at
                                                            times
                                                            $T=\frac{(2k+1)\pi}{2}$,
                                                            $k\in
                                                            Z$.
\section{Photon and excitation transfer between the atom-cavity systems}
\subsection{Photon transition}
Hereafter, we consider the resonance case $\Delta=0$ and take
$\lambda=\frac{1}{\sqrt{2}}$. In order to investigate photon
transition between the cavities, we consider the initial state
$\ket{\psi(0)}=\ket{g,2}\ket{g,0}... \ket{g,0}$, with initial
conditions $A_j(0)=0$, $C_0(0)=1$ and $C_l(0)=0$ for all $j=0,1,...,
n-1$ and $l=1,\ldots, n-1$. Then, one can obtain via equation
(\ref{basis}), $A'_j(0)=0$, and
$C'_j(0)=\frac{1}{\sqrt{n}}\chi_j(\alpha_0)=\frac{d_j}{\sqrt{n}}=\frac{1}{\sqrt{n}}$
for every $j$. Now, with the aid of Eqs. (\ref{eq1'0}) and
(\ref{eq3}), we obtain
$$C_i(t)=\frac{1}{n}\sum_{j}\frac{e^{-i(\omega+\xi x_j)t}}{\sqrt{[1+(\xi x_j)^2]}}[\sqrt{1+(\xi x_j)^2}\cos t\sqrt{1+(\xi x_j)^2}-i\xi x_j\sin t\sqrt{1+(\xi
x_j)^2}]\chi^*_j(\alpha_i),$$
\begin{equation}\label{eq1'}
\hspace{-7cm}A_i(t)=\frac{-i}{n}\sum_{j}\frac{e^{-i(\omega+\xi
x_j)t}}{\sqrt{[1+(\xi x_j)^2]}}\sin t\sqrt{1+(\xi
x_j)^2}\chi^*_j(\alpha_i).
\end{equation}
\\
\textbf{A. Small hopping $\xi\ll$}\\ The above result indicates that
for small hopping $\xi\ll$, we have
$$C_i(t)\simeq\frac{e^{-i\omega t}\cos t}{n}\sum_{j}\chi^*_j(\alpha_i),$$
\begin{equation}\label{eq5}
A_i(t)\simeq \frac{-ie^{-i\omega t}\sin
t}{n}\sum_{j}\chi^*_j(\alpha_i).
\end{equation}
It would be noticed that from the orthogonality relation
$\sum_i\chi^*_j(\alpha_i)\chi_k(\alpha_i)=n\delta_{j,k}$, we have
clearly
$$\sum_i (|C_i(t)|^2+|A_i(t)|^2)=\frac{1}{n^2}\sum_{j,k}\sum_i \chi^*_j(\alpha_i)\chi_k(\alpha_i)=1.$$
The result (\ref{eq5}) indicates that, at times $T=k\pi$, with $k\in
Z$, we have $A_i(T)=0$ and the probability of transfer of two
photons from the first cavity to the $i$-th cavity is given by
$$P_i(T)=|C_i(T)|^2=\frac{1}{n^2}\sum_{j,k}\chi^*_j(\alpha_i)\chi_k(\alpha_i).$$\\
\textbf{B. Large hopping $\xi\gg$}\\
For large hopping strength $\xi\rightarrow \infty$, the result
(\ref{eq1'}) leads to
$$\hspace{-7cm}C_i(t)\simeq \frac{e^{-i\omega t}}{n}\{\cos
t\sum_{j; x_j=0}\chi^*_j(\alpha_i)+ \sum_{j; x_j\neq 0}e^{-2i\xi
tx_j}\chi^*_j(\alpha_i)\},$$
\begin{equation}\label{eq1'''}
A_i(t)\simeq\frac{-ie^{-i\omega t}}{n}\{\sin t\sum_{j;
x_j=0}\chi^*_j(\alpha_i)+ \sum_{j; x_j\neq 0}\frac{e^{-i\xi
tx_j}\sin (\xi t x_j)}{\xi x_j}\chi^*_j(\alpha_i)\}\simeq
\frac{-ie^{-i\omega t}}{n}\sin t\sum_{j; x_j=0}\chi^*_j(\alpha_i).
\end{equation}
Again, at times $T=k\pi$, with $k\in Z$, we have $A_i(T)\simeq0$ and
the initially unentangled atoms remain unentangled.
\subsection{Excitation transition}
The excitation transfer between the atoms, can be achieved by
preparing the system initially at the state
$\ket{\psi(0)}=\ket{e,0}\ket{g,0}... \ket{g,0}$, with initial
conditions $C_j(0)=0$, $A_0(0)=1$ and $A_l(0)=0$ for all $j=0,1,...,
n-1$ and $l=1,\ldots, n-1$. Then, one can obtain via equation
(\ref{basis}), $C'_j(0)=0$, and $A'_j(0)=\frac{1}{\sqrt{n}}$ for
every $j$. Again, with the aid of Eqs. (\ref{eq1'0}) and
(\ref{eq3}), one can obtain
$$\hspace{-6cm} C_i(t)=\frac{-i}{n}\sum_{j}\frac{e^{-i(\omega+\xi
x_j)t}}{\sqrt{[1+(\xi x_j)^2]}}\sin t\sqrt{1+(\xi
x_j)^2}\chi^*_j(\alpha_i),$$
\begin{equation}\label{eq10'}
A_i(t)=\frac{1}{n}\sum_{j}\frac{e^{-i(\omega+\xi
x_j)t}}{\sqrt{[1+(\xi x_j)^2]}}[\sqrt{1+(\xi x_j)^2}\cos
t\sqrt{1+(\xi x_j)^2}+i\xi x_j\sin t\sqrt{1+(\xi
x_j)^2}]\chi^*_j(\alpha_i).
\end{equation}
Similar to the photon transition (as argued in the previous
subsection), one can discuss excitation transfer for small and large
hopping limits, where it is seen that at the limiting cases $\xi\ll$
and $\xi\gg$, we have $C_i(T)\simeq 0$  for times $T=l\pi$, with
$l\in Z$; The probability of transfer of atomic excitation from the
first atom to the $i$-th atom, i.e. $P_i(T)=|A_i(T)|^2$, can be
evaluated similarly.
\section{Examples}
In this section we give two examples of uncolored Cayley networks
for which the photonic and atomic state transfer
between atom-cavities is investigated in details.\\
\textbf{1. The Cycle network $C_n$}\\
 If $G=Z_n$  is the finite cyclic group of order n and the set $S$ consists of two elements, the standard generator of $G$ and its inverse, i.e., $a,a^{-1}$ with $a^n=1$, then the Cayley graph is the cycle
 $C_n$. In this case the irreducible characters of the group are given by $\chi_k(a^l)=\omega^{kl}$, for $k,l=0,1,\ldots, n-1$, where $\omega:=e^{\frac{2\pi i}{n}}$ is the $n$-th root of identity.\\
The adjacency matrix of the network is $A=S+S^{-1}$, where $S$ is
the shift matrix of order $n$ so that we have $S^n=I$. Then, the
corresponding eigenvalues of $A$ are given by
$x_l=\chi_l(a)+\chi_l(a^{-1})=\omega^{l}+\omega^{-l}=2\cos
(\frac{2\pi l}{n})$, $l=0,1,\ldots, n-1$. Now, one can evaluate the
probability amplitudes corresponding to the excitation or photon
transition via the results (\ref{eq1'}). For instance, in the case
of even $n=2m$ with odd $m=2l+1$, we have $x_j=\cos \frac{\pi
j}{m}\neq 0$ for $j=0,1,\ldots, 2m-1$, so that
$x_j=x_{2m-j}=-x_{m-j}$. Then, for the photon transition process,
the results (\ref{eq5}) and (\ref{eq1'''}) lead to
$$C_i(t)\simeq\frac{e^{-i\omega t}\cos t}{2m}\sum_{j}\omega^{-ij}=e^{-i\omega t}\cos t\;\ \delta_{i,0},$$
\begin{equation}\label{eq5}
A_i(t)\simeq \frac{-ie^{-i\omega t}\sin
t}{2m}\sum_{j}\omega^{-ij}=e^{-i\omega t}\sin t\;\ \delta_{i,0}
\end{equation}
and
$$C_i(t)\simeq \frac{e^{-i\omega t}}{2m}\sum_{j=0}^{2m-1}e^{-2i\xi
tx_j}\omega^{-ij}=\frac{e^{-i\omega t}}{2m}\{e^{-4i\xi
t}+(-1)^ie^{4i\xi t}+2\sum_{j=1}^{m-1}e^{-2i\xi x_j t}\cos \frac{\pi
ji}{m}\},$$
\begin{equation}\label{eq1'''}
A_i(t)\simeq\frac{-ie^{-i\omega t}}{2m}\sum_{j}\frac{e^{-i\xi
tx_j}\sin (\xi t x_j)}{\xi x_j}\omega^{-ij}\simeq 0
\end{equation}
for the asymptotic limits $\xi\ll$ and $\xi\gg$, respectively. The
above relations indicate that for small hopping $\xi\rightarrow 0$,
we have
$$C_0(t)\simeq e^{-i\omega t} \cos t,\;\;\ A_0(t)\simeq e^{-i\omega t} \sin t,\;\;\;\ C_i(t)=A_i(t)=0; \;\ \mbox{for}\;\ i\neq 0,$$
so that after time $t$, the initial state
$\ket{\psi(0)}=\ket{g,2}\ket{g,0}\ldots \ket{g,0}$ evolves as
$$\ket{\psi(t)}\simeq e^{-i\omega t}(\cos t \ket{g,2}+\sin t\ket{e,0})\ket{g,0}\ldots \ket{g,0}.$$
Therefore, the photon transition can not be achieved and a quantum
correlation is generated between the atomic and photonic states of
the first cavity. On the other hand, for large hopping
$\xi\rightarrow \infty$, one obtains $A_i(t)\simeq 0$ for all times
and
$$C_{i=odd}(t)\simeq \frac{-ie^{-i\omega t}}{m}\{\sin 4\xi t +2\sum_{j=1}^{\frac{m-1}{2}}\sin (2\xi x_j t)\cos \frac{\pi
ji}{m}\},$$
$$C_{i=even}(t)\simeq \frac{e^{-i\omega t}}{m}\{\cos 4\xi t +2\sum_{j=1}^{\frac{m-1}{2}}\cos (2\xi x_j t)\cos \frac{\pi
ji}{m}\}.$$ For example, for $n=6$ with $m=3$, we have $x_0=2$,
$x_1=x_5=1$, $x_2=x_4=-1$, $x_3=-2$ and the above amplitudes are
given by
$$C_0(t)=\frac{e^{-i\omega t}}{3}\{\cos 4\xi t +2\cos 2\xi t\},$$
$$C_1(t)=C_5(t)=\frac{-ie^{-i\omega t}}{3}\{\sin 4\xi t +\sin 2\xi t\},$$
$$C_2(t)=C_4(t)=\frac{-ie^{-i\omega t}}{3}\{\cos 4\xi t -\cos 2\xi t\},$$
$$C_3(t)=\frac{-ie^{-i\omega t}}{3}\{\sin 4\xi t -2\sin 2\xi t\},$$
Then, after time $T=\frac{3\pi}{8\xi}$, we obtain the probabilities
associated with two photon transition from the first cavity to the
other cavities as follows
$$P_0(T)=\frac{2}{9}\simeq 0.22,\;\;\  P_1(T)=P_5(T)=0.01,\;\;\ P_2(T)=P_4(T)=0.05,\;\;\ P_3(T)=\frac{5.76}{9}\simeq 0.64.$$
\textbf{2. The Hypercube network $Q_d$} \\
The Cayley graph of the direct product of groups $Z_2$, $d$ times
(with the cartesian product of generating sets as a generating set)
is the cartesian product of the corresponding Cayley graphs, which
gives the hypercube graph $Q_d$ (called also the Hamming network
$H(d,2)$). Thus, the Cayley graph of the abelian group
$G=\underbrace{Z_2\times Z_2\times\ldots\times Z_2}_{d- times}$ with
the set of generators consisting of $n$ elements $(1,0,0,\ldots,0),
(0,1,0,\ldots, 0),\ldots, (0,0,\ldots,0,1)$, is the hypercube
network $Q_d$ with $n=2^d$ nodes and the adjacency matrix
$$A=\sum_{i=1}^d I\otimes \ldots \otimes I\otimes \underbrace{\sigma_x}_{i-th}\otimes I \ldots \otimes
I.$$ The irreducible characters of the group $G$ are direct products
of the irreducible characters of the group $Z_2$, i.e., by denoting
$k,l\in G$ in the binary representation $k\equiv k_{d-1}\ldots
k_1k_0$ and $l\equiv l_{d-1}\ldots l_1l_0$, with $k_i, l_i\in
\{0,1\}$ (so that $k=k_0+2^1k_1+\ldots + 2^{d-1}k_{d-1}$), we have
$\chi_k(l)=(-1)^{k\cdot l}$, where $k\cdot
l=\sum_{i=0}^{d-1}k_il_i$. Also, one can easily show that the
eigenvalues of the adjacency matrix $A$ with the corresponding
degeneracy degrees are given by $x_i=d-2i$ and
$D(x_i)=\frac{d!}{i!(d-i)!}$  for $i=0,1,\ldots, d$, respectively.

For instance, for $d=2$, we have $x_0=2$, $x_1=x_2=0$ and $x_3=-2$.
Then, the Eq. (\ref{eq1'}) leads to
$$C_i(t)=\frac{e^{-i\omega t}}{4\sqrt{1+4\xi^2}}\{\sqrt{1+4\xi^2}(e^{-2i\xi t}+\chi_3(i)e^{2i\xi t})\cos t\sqrt{1+4\xi^2}-2i\xi (e^{-2i\xi t}-\chi_3(i)e^{2i\xi t})\sin t\sqrt{1+4\xi^2}+$$
$$\sqrt{1+4\xi^2}(\chi_1(i)+\chi_2(i))\cos t\},$$
$$A_i(t)=\frac{-ie^{-i\omega t}}{4\sqrt{1+4\xi^2}}\{(e^{-2i\xi t}+\chi_3(i)e^{2i\xi t})\sin t\sqrt{1+4\xi^2}+\sqrt{1+4\xi^2}(\chi_1(i)+\chi_2(i))\sin t\}.$$
The above results are respectively simplified for the small and
large hopping parameter $\xi$, as follows
$$\hspace{-2cm}C_i(t)\simeq\frac{e^{-i\omega t}\cos t}{4}\{e^{-2i\xi t}+\chi_3(i)e^{2i\xi t}+\chi_1(i)+\chi_2(i)\},$$
$$A_i(t)\simeq\frac{-ie^{-i\omega t}\sin t}{4}\{e^{-2i\xi t}+\chi_3(i)e^{2i\xi t}+\chi_1(i)+\chi_2(i)\};\;\;\;\;\ \mbox{for}\;\;\ \xi\ll$$
and
$$C_i(t)\simeq\frac{e^{-i\omega t}}{4}\{e^{-4i\xi t}+\chi_3(i)e^{4i\xi t}+(\chi_1(i)+\chi_2(i))\cos t\},$$
$$A_i(t)\simeq \frac{-ie^{-i\omega t}}{4}(\chi_1(i)+\chi_2(i))\sin t,\;\;\;\;\;\ \mbox{for}\;\;\ \xi\gg.$$
By substituting $\chi_1(0)=\chi_1(2)=1, \;\ \chi_1(1)=\chi_1(3)=-1$,
$\chi_2(0)=\chi_2(1)=1, \;\ \chi_2(2)=\chi_2(3)=-1$ and
$\chi_3(0)=\chi_3(3)=1, \;\ \chi_3(1)=\chi_3(2)=-1$ in the above
relations, the corresponding amplitudes are determined for every
$i=0,1,2,3$.
\section{Conclusion}
In summery, quantum state transfer and entanglement generation
between $n$ coupled atom-cavity systems with uncolored Cayley
interacting networks, was analyzed. By employing the
excitation-photon conservation symmetry of the Hamiltonian and
introducing some suitable `generalized' Fourier transformed basis
states for the state space of the system, the corresponding Hilbert
space was block-diagonalized with $2$ dimensional blocks. Due to
this useful reduction, the corresponding Shr\"{o}dinger equation was
solved exactly, and state transfer (excitation or photon transition
between the atoms or the cavities) and entanglement generation
between the atoms were discussed. The large and small hopping limits
was discussed where, it was shown that for large hopping strength,
the initially unentangled atoms remain effectively unentangled
forever whereas for small hopping, the initially unentangled atoms
become entangled and the initial photonic $W$-state oscillates
between atomic and photonic $W$-states periodically.


\begin{thebibliography}{99}
\bibitem{11} M. Christandl, N. Datta, A.
Ekert, and A. J. Landahl, Phys. Rev. Lett. 92, 187902, (2004).
\bibitem{12}
M. Christandl, N. Datta, T. C. Dorlas, A. Ekert, A. Kay and A. J.
Landahl, Phys. Rev. A 71, 032312, (2005).
\bibitem{13}C. Facer, J. Twamley and J. Cresser,
Phys. Rev. A 77, 012334, (2008).
\bibitem{14}
D. Burgarth and S. Bose, Phys. Rev. A 71, 052315, (2005).
\bibitem{15}
D. Burgarth and S. Bose, New J. Phys. 7, 135, (2005).
\bibitem{16} M. H. Yung and S. Bose, Phys. Rev. A 71,
032310, (2005).
\bibitem{17} M. H. Yung, Phys. Rev. A 74, 030303, (2006).
\bibitem{18}M. A. Jafarizadeh and R. Sufiani, Phys. Rev. A 77,
022315, (2008).
\bibitem{19}M. A. Jafarizadeh, et al., J. Phys. A: Math. Theor. 41,
475302, (2008).
\bibitem{20}
M A Jafarizadeh, et al., J. Stat. Mech. 05014, (2011).
\bibitem{ref5}
Turchette, Q.A., Hood, C.J., Lange, W., Mabuchi, H., Kimble, H.J.,
Phys. Rev. Lett. 75, 4710 (1995); M. Brune et al., Phys. Rev. Lett.
77, 4887 (1996).
\bibitem{ref6}
Mattle, K., Weinfurter, H., Kwiat, P.G., Zeilinger, A., Phys. Rev.
Lett. 76, 4656 (1996).
\bibitem{Alex1}
M. Alexanian, Phys. Rev. A 83, 023814, 2011.
\bibitem{ref1}
A. Biswas and G. S. Agarwal, Phys. Rev. A 70, 022323 (2004).
\bibitem{ref}
Cirac,J.I., Zoller,P., Kimble,H.J., Mabuchi,H., Phys. Rev. Lett. 78,
3221, (1997).
\bibitem{Alex2}
M. Alexanian, Generation of entangled tripartite states in three
identical cavities, arxiv: quant-ph/12034173.
\bibitem{ozgur}
A.\"{U}.C. Hardal, \"{O}.E. Müstecaplioglu, J. Opt. Soc. Am. B 29,
1822-1828, (2012).
\bibitem{suf}
R. Sufiani and A. Darkhosh, Generation of quantum entanglement
between three level atoms via $n$ coupled cavities, arxiv:
quant-ph/13033861.
\bibitem{book1}
R. R. Puri, \it{{Mathematical methods of quantum optics}}, Springer,
(2001).
\bibitem{book2}
M. O. Scully and M. S. Zubairy, \it{{Quantum Optics}}, Cambridge
University Press, (1997).
\bibitem{Gordon}
Gordon James, Martin Liebeck,(1993), {\it Representations and
characters of groups} ( Cambridge University Press, Cambridge).
\bibitem{JC}
Andrei B. Klimov and Sergei M. Chumakov, (2009), {\it A
Group-Theoretical Approach to Quantum Optics} (WILEY-VCH Verlag GmbH
and Co. KGaA, Weinheim)
\bibitem{peres1}
A. Peres, Phys. Rev. Lett, 77, pp. 1413-1415, (1996).
\bibitem{peres2}
M. Horodecki, et. al, Phys. Lett. A, 223, pp. 1-8, (1996).
\end{thebibliography}
\end{document}